# In-plane Chiral Tunneling and Out-of-plane Valley-polarized Quantum Tunneling in Twisted Graphene Trilayer

Jia-Bin Qiao and Lin He*

Department of Physics, Beijing Normal University, Beijing, 100875, People's Republic of China

**Here we show that twisted graphene trilayer made by misoriented stacking a graphene monolayer on top of a Bernal graphene bilayer can exhibit rich and tailored electronic properties. For the case that the graphene monolayer and bilayer are strongly coupled, both the massless Dirac fermions and massive chiral fermions coexist in the twisted trilayer, leading to unique in-plane chiral tunneling. For a weak coupling between the two graphene systems, the distinct chiralities and pseudospin textures of quasiparticles in monolayer and bilayer enable vertical valley-polarized quantum tunneling between them. Intriguingly, the polarity of the valley polarization can be inverted simply by either controlling the rotational angles between the two systems or tuning the Fermi levels of the two systems. Our result implies that layered van der Waals structures assembled from individual atomic planes can create materials that harbor unusual properties and new functionalities depending on the stacking configuration of the crystalline layers.**

## I. INTRODUCTION

Very recently, layered van der Waals crystals made by stacking two-dimensional (2D) atomic crystals on top of each other have attracted much attention [1]. These crystals engineered with one-atomic-plane accuracy provide unprecedented opportunities to explore unusual properties and new phenomena [2-15]. It was demonstrated that properties of the van der Waals structures depend not only on their building blocks with layer-specific attributes, but also on how the 2D atomic crystals are stacked [12,16-19]. A remarkable van der Waals crystal revealing stacking-sensitive properties is graphene bilayer. In Bernal stacking, the charge carriers of the graphene bilayer have a parabolic energy spectrum and exhibit chirality that resembles those associated with spin 1. However, a twist between the two layers splits the parabolic band touching into two Dirac cones [13-16,19-24] and changes the chirality of the low-energy quasiparticles to those of spin 1/2 [18,25]. This provides a facile route to tune the electronic properties of graphene bilayer.

In this paper, we show that twisted graphene trilayer made by misoriented stacking a graphene monolayer on top of a Bernal graphene bilayer can exhibit tailored electronic properties. For a strong coupling between the monolayer and the bilayer, the twisted graphene trilayer combines the properties of the building blocks, *i.e.*, both the massless Dirac fermions and massive chiral fermions coexist, revealing unique and tunable in-plane chiral tunneling. For a weak coupling between the two graphene systems, we demonstrate that it is possible to realize out-of-plane valley-polarized quantum tunneling between the two graphene systems and, importantly, the polarity of the valley polarization can be simply inverted by either controlling the rotational angles or tunning the Fermi levels of the two graphene systems.

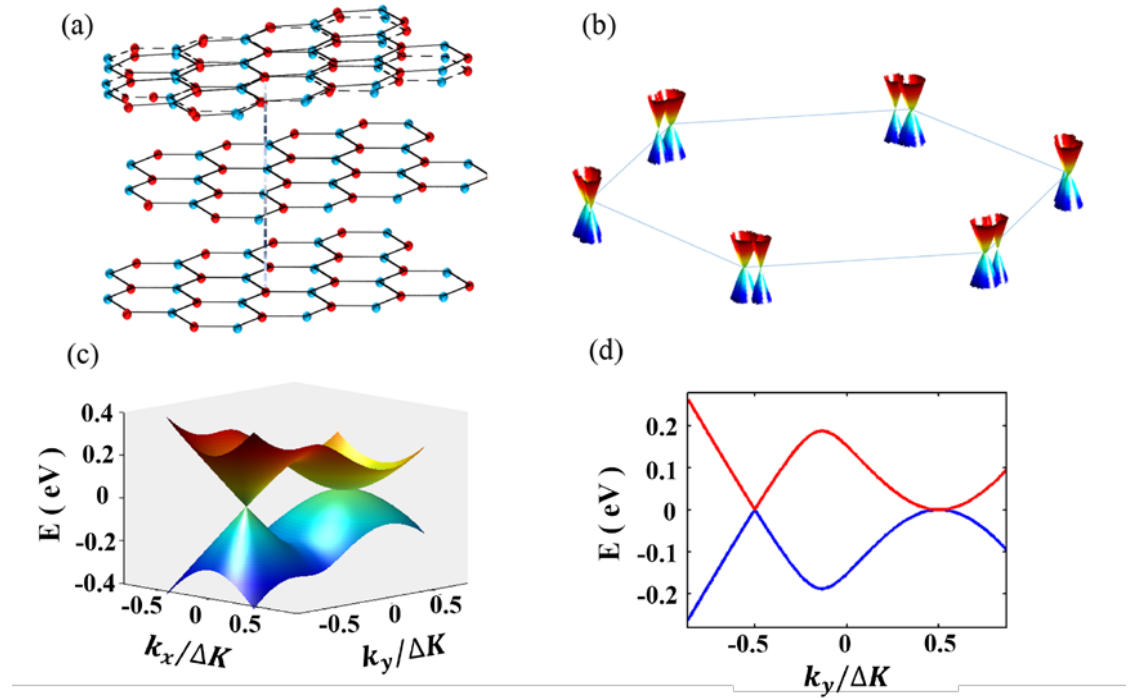

**FIG. 1. (color online)** (a) Schematic representation of graphene trilayer. The red and blue balls denote the two A and B sublattices, respectively. The initial structure of the graphene trilayer is the ABA stacking. A stacking misorientation of the top layer results in a twisted graphene trilayer. (b) The schematic band structures of the twisted graphene trilayer. The separation of the Dirac cones and the parabolic valleys is attributed to the rotations between the top monolayer and the underlaying bilayer. (c) The low-energy electronic spectrum of a twisted graphene trilayer with a rotational angle of 3.89°. (d) A section view along $k_x = 0$ of the electronic band structure in panel (c).

## II. IN-PLANE CHIRAL TUNNELING

Figure 1(a) shows a schematic structure of a twisted graphene trilayer composed of graphene bilayer with Bernal stacking and a third one with a relative rotation angle $\theta$ with respect to the other two from an initial ABA stacking. The stacking misorientation of the third layer leads to the appearance of Moiré patterns in the structure, which have been observed in recent experiments [26-29]. More importantly, the rotation of the third layer separates the Dirac cones of the twisted monolayer, $K_S$ (or $K'_S$), and the parabolic bands of the Bernal bilayer, $K_B$ (or $K'_B$), in the reciprocal space, as shown in Fig. 1(b). Here $K$ and $K'$ are the valley-isospin degree of freedom in graphene systems. The relative shift between $K_S$ and $K_B$ is $\Delta K = (\Delta K_x, \Delta K_y) = 2|\vec{K}|\sin(\theta/2)$ with $|\vec{K}| = 4\pi/3a$ and $a \sim 0.246$ nm the lattice constant of the hexagonal lattice [13,25]. The displaced Dirac cones and parabolic valleys cross at the intersections and two saddle points (van Hove singularities) at $\pm E_V$ emerge in the presence of a finite interlayer coupling [29,30], as shown in Fig. 1(c) and 1(d). For simplicity, we set $\Delta K_x = 0$ and $\Delta K_y = \Delta K$ in Fig. 1 (See supporting materials [31] for details of calculation). Such a band structure of the twisted trilayer was demonstrated very recently by the observation of two low-energy van Hove singularities [29].

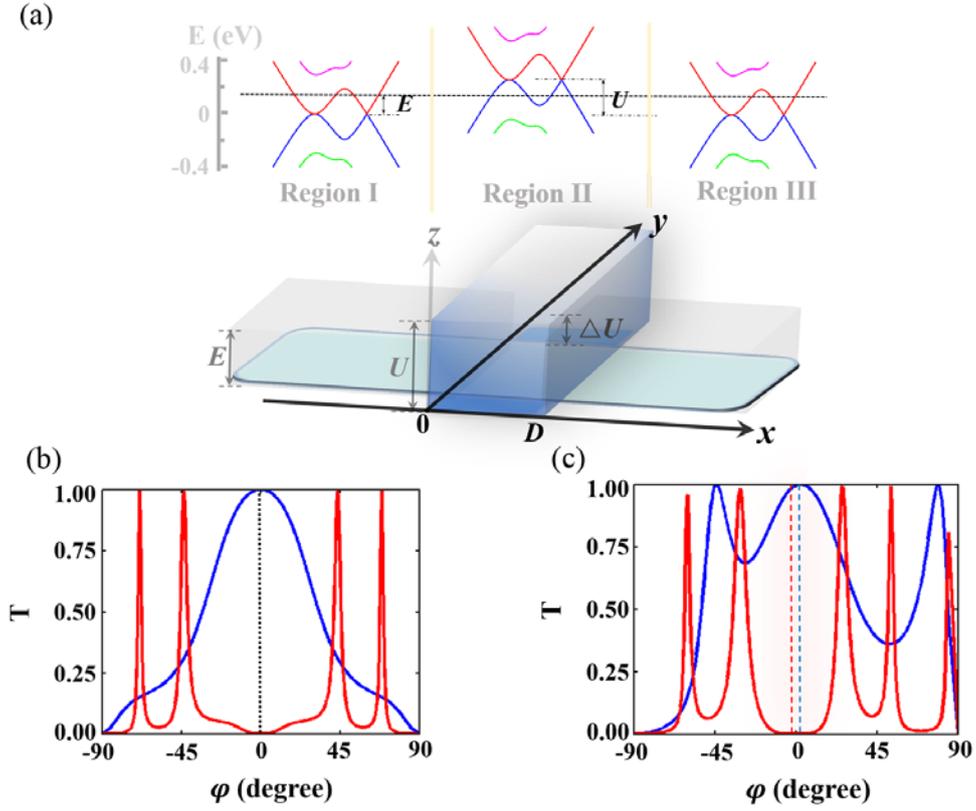

**FIG. 2. (color online)** In-plane chiral tunneling in a twisted graphene trilayer. (a) Schematic diagrams of the band spectrum and the positions of the Fermi energy $E$ of a twisted trilayer in the three regions of a tunneling process. The red and blue curves denote the low-energy bands of the twisted graphene trilayer, while the pink and green curves are the next bands of the system. The height of the potential barrier is $U = E + \Delta U$ and the width is $D$. (b) Transmission probability through a 100 nm-wide barrier as a function of the incident angle for single-layer (blue curve) and Bernal bi-layer (red curve) graphene. The two curves are symmetric about the axis $\varphi = 0°$. (c) Transmission probability through a 100 nm-wide barrier as a function of the incident angle for quasiparticles in the sub-valley $K_S$ (blue curve) and $K_B$ (red curve) of a twisted trilayer with $\theta = 3.89°$. Other parameters used in the calculation of panels (b) and (c) are $E = 0.05$ eV and $\Delta U = 0.10$ eV.

Obviously, both the massless Dirac fermions and massive chiral fermions coexist in the twisted graphene trilayer, which implies that the trilayer may combine the properties of the twisted monolayer and the Bernal bilayer. In order to elucidate the similarities and differences of the three graphene systems (monolayer, Bernal bilayer, and the twisted trilayer), we calculated chiral tunneling [25,32] of these systems for comparison. Figure 2(a) shows the general scheme for the chiral tunneling of quasiparticles in twisted trilayer through an infinite potential barrier along the $y$ axis $U(x)$, which has a rectangular shape with width $D$. The incident energy of the electron is $E$ and the energy difference between the potential barrier and the electron is $\Delta U$. The rectangular shape assumption of the barrier [25,32] forbids electron scattering to mix the two valleys, $K$ and $K'$, in graphene and allows us to consider scattering electrons within one valley. If we know the wave functions in the three regions of the tunneling process, *i.e.*, the left of the barrier ($x < 0$), inside the barrier ($0 < x < D$), and the right of the barrier ($x > D$), then the transmission coefficient can be determined from the continuity of the wave functions and their derivatives. Figure 2(b) shows example of the transmission probability as a function of the incident angle $T(\varphi)$ for a graphene monolayer and a Bernal bilayer. For electrons incident in the normal direction $\varphi = 0$, the electrons are transformed to propagating (exponential decaying) holes inside the barrier resulting in perfect transmission (reflection) in a graphene monolayer (bilayer). The two distinct behaviors, *i.e.*, the perfect tunneling in graphene monolayer and the perfect reflection in Bernal bilayer, are viewed as two incarnations of the Klein paradox [32].

For the twisted graphene trilayer, the low-energy Hamiltonian is described by [29]

$$H = \begin{pmatrix} H_1(k - \Delta K/2) & H_\perp^{(12)} & 0 \\ H_\perp^{(21)} & H_2(k + \Delta K/2) & H_\perp^{(23)} \\ 0 & H_\perp^{(32)} & H_3(k + \Delta K/2) \end{pmatrix}. \quad (1)$$

Where $H_\tau(k) \equiv v_F \begin{pmatrix} 0 & k^* \\ k & 0 \end{pmatrix}$ with $\tau$ ($\tau = 1,2,3$) enumerating the layers is the Hamiltonian of single-layer graphene (here $v_F$ is the Fermi velocity and $k = k_x + ik_y$ is the wave vector), and $H_\perp$ denotes the interlayer coupling between different layers. The wave function of the twisted trilayer is the vector $\Phi = (\psi_1^A, \psi_1^B, \psi_2^A, \psi_2^B, \psi_3^A, \psi_3^B)^+$, where the subscript numbers enumerate the layers and the superscript $A$ and $B$ denote sublattices on each layer. Figure 2(c) shows the transmission probability of quasiparticles in the sub-valleys, $K_S$ and $K_B$, of the twisted trilayer as a function of the incident angle $T(\varphi)$ (See supporting materials [31] for details of calculation). For comparison, we use the same parameters, $E$, $\Delta U$, and $D$, in the calculation of Fig. 2(b) and Fig. 2(c).

The main characteristics of $T(\varphi)$ in the $K_S$ and $K_B$ resemble that in graphene monolayer and Bernal bilayer, respectively. An obvious difference is that $T(\varphi)$ is asymmetric about $\varphi = 0$ in the twisted graphene trilayer. Similar behavior is also obtained in twisted graphene bilayer [25]. The origin of this asymmetry arises from the fact that the quasiparticles in the $K_S$ ($K_B$) are not absolute massless Dirac fermions as in graphene monolayer (massive chiral fermions as in Bernal bilayer). The wavefunction of the quasiparticle in twisted graphene trilayer is a function of the energy. In the twisted trilayer, we also can apply an external electric field perpendicular to the system to open a gap in the sub-valley $K_B$, which ensures that only quasiparticles in the $K_S$ contribute to the tunneling process.

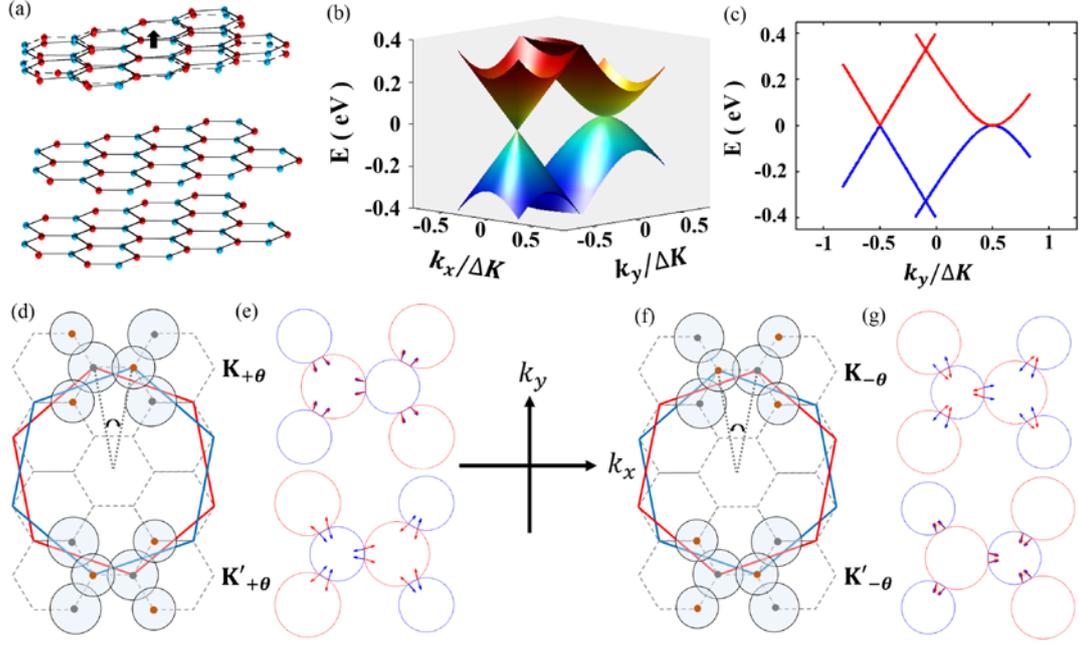

**FIG. 3. (color online)** (a) Schematic structure of a twisted trilayer with a weak coupling between the twisted monolayer and the Bernal bilayer. The distance between the twisted layer and the Bernal bilayer can be ranging from 1 nm to several nanometers. (b) The electronic spectrum of quasiparticles in the trilayer of panel (a). (c) The section of the electronic spectrum in panel (b) along $k_x = 0$. Brillouin zone of the trilayer in panel (a) with (d) a clockwise rotational angle and (f) a counter-clockwise rotational angle. The blue (red) solid large hexagons correspond to the first Brillouin zone of the monolayer and the Bernal bilayer, respectively. The twisted angle is defined as positive (negative) for clockwise (counter-clockwise) rotation of the monolayer. The black dashed hexagons correspond to the reduced Brillouin zone of the trilayer. (e) and (g) show the pseudo-spin textures around the $K$ and $K'$ valleys of graphene monolayer (indicated by the blue circle and the pseudo-spin texture is denoted by the blue arrows) and Bernal bilayer (indicated by the red circle and the pseudo-spin texture is denoted by the red arrows) at the intersection points of their Fermi circles with $E \sim 0.3$ eV. In panel (e), the rotational angle is positive, while it is negative in panel (g).

## III. OUT-OF-PLANE VALLEY-POLARIZED QUANTUM TUNNELING

Now we begin to consider the out-of-plane quantum tunneling in twisted graphene trilayer with a weak interlayer coupling between the twisted layer and the Bernal bilayer. Here we assume that the interlayer coupling is too weak to generate the low-energy saddle points in the trilayer, as shown in Fig. 1, however, the monolayer and the bilayer are not isolated each other to ensure the out-of-plane quantum tunneling between them (the interlayer distance is ranging from 1 nm to several nanometers). Such a structure (Fig. 3(a)) can be realized by inserting a thin insulator, such as thin hexagonal boron nitride, between the two graphene systems, as demonstrated in experiments very recently [2,9,17]. Theoretical descriptions of the quantum tunneling between parallel 2D conductors have been studied with great success for many years [33-37], and the Hamiltonian to describe the tunneling system in Fig. 3(a) should be

$$\boldsymbol{H} = \begin{pmatrix} H_1(k - \Delta K/2) & \gamma^+ \\ \gamma & H_{eff}^{(B)} \end{pmatrix}. \qquad (2)$$

Here, $H_{eff}^{(B)}$ represents the effective Hamiltonian of the Bernal graphene bilayer and $\gamma$ is the transition matrix depicting the tunnel coupling between the twisted monolayer and the Bernal bilayer [35]. Figure 3(b) and 3(c) show the electronic band structure of the system in the first Brillouin zone of the twisted monolayer and Bernal bilayer (See supporting materials [31] for details of calculation). Because of the periodic lattice structure of the twisted trilayer, each $K_S$ ($K'_S$) of monolayer has three adjacent $K_B$ ($K'_B$) of bilayer and, similarly, each $K_B$ ($K'_B$) of bilayer has three adjacent $K_S$ ($K'_S$) of monolayer, as shown in Fig. 3(d) and Fig. 3(f). Similar result has also been obtained in twisted graphene bilayer [38-41]. Therefore, each monolayer Fermi circle intersects with three adjacent Fermi circles of bilayer, as shown in Fig. 3(e) and Fig. 3(g). For simplicity, we only take into account nondissipative quantum tunneling between the two graphene systems, which occurs at the intersection points of their energy spectra, and the pseudo-spin is also assumed to be conserved in a tunneling event.

Intriguingly, our analysis indicates that the pseudo-spin textures at the intersection points of the Fermi circles are rather different between the two valleys $K$ and $K'$, as shown in Fig. 3(d-g). The ability to generate valley-asymmetric properties plays a vital role in realizing the valley-based spintronics, the so-called valleytronics, in graphene and graphene-like systems [42-49]. Therefore, the valley-asymmetric pseudo-spin textures, as shown in Fig. 3(e) and 3(g), imply the possibility to realize valley-polarized quantum tunneling between the two graphene systems. Here we point out that although each Fermi circle of monolayer intersects with three adjacent Fermi circles of Bernal bilayer, as shown in Fig. 3(e) and 3(g), the pseudo-spin textures at the intersection points of every adjacent Fermi circles of the two graphene systems are the same. Therefore, it is reasonable to calculate the interlayer valley polarization only using two intersection Fermi circles, i.e., a $K_B$ ($K'_B$) valley of Bernal bilayer and a $K_S$ ($K'_S$) of monolayer. Additionally, at the intersection points, the relative angles between the pseudo-spin of the monolayer and the bilayer in the two valleys $K$ and $K'$ are mirror symmetric about $\theta = 0°$, as shown in Fig. 3(e) and 3(g). It indicates that if we obtain 100% valley-polarized tunneling of the $K$ valley at a positive rotational angle $+\theta$ [here the angle is defined as positive (negative) for clockwise (counter-clockwise) rotation of the graphene monolayer], then we could generate out-of-plane tunneling current fully from the $K'$ valley at $-\theta$.

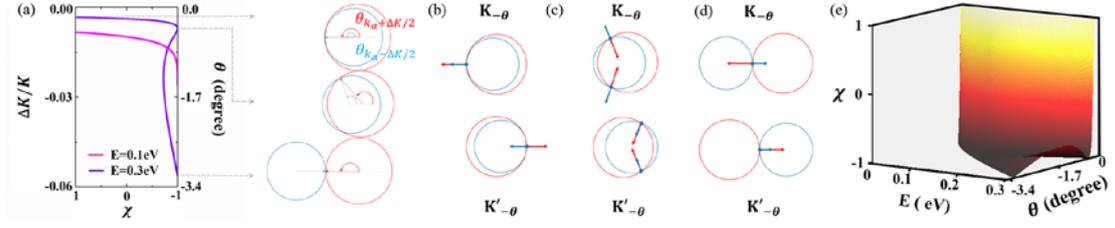

**FIG. 4. (color online)** (a) The valley polarization $\chi$ as a function of the rotational angle (or momentum shift, as shown in left-Y axis). The pink and purple curves correspond to $E = 0.1$ eV and $E = 0.3$ eV, respectively. The left panel shows three typical structures of the Fermi circles in graphene monolayer (blue circles) and bilayer (red circles) that result in fully valley-polarized tunneling for $E = 0.3$ eV. (b), (c), and (d) show the pseudo-spin textures at the intersection points of the two Fermi circles for the three valley-polarized structures in panel (a). (e) The valley polarization as a function of the rotational angle and the Fermi energy.

To verify the above analysis, we calculate the valley polarization of the out-of-plane tunneling between the twisted monolayer and the Bernal bilayer

$$\chi = \frac{G^{(K)} - G^{(K')}}{G^{(K)} + G^{(K')}}. \qquad (3)$$

Here, $G^{(K)}$ ($G^{(K')}$) is the linear magneto-tunneling conductance for electrons from valley $K$ ($K'$) of graphene. $\chi = 1$ indicates 100% valley-polarized tunneling current from the valley $K$ ($G^{(K')} = 0$), while $\chi = -1$ means fully valley polarization of the tunneling current from the valley $K'$ ($G^{(K)} = 0$). According to our analysis, the valley polarization $\chi$ only depends on the pseudo-spin matrix elements, which describe overlap elements between the spinors at the two intersection points of the Fermi circles (See supporting materials [31] for details). For the quantum tunneling with pseudo-spin conservation, Eq. (3) can be simplified to

$$\chi(\theta) = -\cos\left(2\theta_{k_\alpha + \Delta K/2} + \theta_{k_\alpha - \Delta K/2}\right), \qquad (4)$$

where $\theta_{k_\alpha \pm \Delta K/2}$ denotes the angle of momentum wave vector $k_\alpha \pm \Delta K/2$ with the positive $k_x$ axis.

Figure 4(a) shows two $\chi(\theta)$ curves as a function of counter-clockwise rotation of the graphene monolayer (*i.e.*, for negative twisted angle) with different Fermi energies, $E = 0.1$ eV and $E = 0.3$ eV. For positive twisted angle, we could obtain similar result but with $\chi(-\theta) = -\chi(\theta)$. For $\theta = 0^\circ$, the two Fermi circles of graphene monolayer and bilayer are concentric and there is no intersection point of the Fermi circles because of that $k_F^{(B)} > k_F^{(S)}$ at low energy (here $k_F^{(B)}$ and $k_F^{(S)}$ are the Fermi wave vectors of graphene bilayer and single-layer respectively). Therefore, the out-of-plane tunneling is forbidden when $-(k_F^{(B)} - k_F^{(S)}) < \Delta K < 0$, as shown in Fig. 4(a). For $\Delta K = -(k_F^{(B)} - k_F^{(S)})$, the Fermi circle of graphene monolayer becomes inscribed circle of that of Bernal bilayer. At the intersection points, the pseudo-spins of the two graphene systems are parallel in the valley *K* and anti-parallel in the valley *K'*, as shown in Fig. 4(b). Consequently, we obtain $\chi = 1$. For $\Delta K = -(k_F^{(B)} + k_F^{(S)})$, the Fermi circle of graphene monolayer becomes excircle of that of Bernal bilayer, as shown in Fig. 4(d). Then, we obtain $\chi = -1$ because of that the pseudo-spins of the two graphene systems are parallel in the valley *K'* and anti-parallel in the valley *K*. At small Fermi energy ($E < 0.15$ eV), the valley polarization changes monotonously from 1 to -1 with increasing the twisted angle when $\Delta K \in [-(k_F^{(B)} + k_F^{(S)}), -(k_F^{(B)} - k_F^{(S)})]$, as shown in Fig. 4(a) and Fig. 4(e). However, at large Fermi energy ($E > 0.15$ eV), for example $E = 0.3$ eV, there is a "magic" twisted angle within $-(k_F^{(B)} + k_F^{(S)}) < \Delta K < -(k_F^{(B)} - k_F^{(S)})$ that ensures $2\theta_{k_a + \Delta K/2} + \theta_{k_a - \Delta K/2} = 2\pi$ (as shown in Fig. 4(c)) and results in $\chi = -1$ (as shown in Fig. 4(a) and Fig. 4(e)). Obviously, the result in Fig. 4 indicates that it is possible to realize valley-polarized quantum tunneling between the two graphene systems and the polarity of the valley polarization can be inverted by controlling the rotational angles. For a twisted trilayer with a fixed rotation angle, it is more facile to tune the valley polarization from 1 to -1, or vice versa, by changing the Fermi levels of the two graphene systems, as shown in Fig. 4(e). With steady improvement in fabrication techniques of van der Waals heterostructures [1], such an artificial structure, as shown in Fig. 3(a), is within the grasp of today's technology and the Fermi level is easily tuned by the electric field effect. Therefore, the predicted effect of this paper is expected to be realized in the very near future.

## IV. CONCLUSION

In summary, we demonstrate that it is facile to tailor the electronic properties of twisted graphene trilayer. New functionality, such as valley-polarized tunneling, can be realized by adjusting the stacking arrangement of the twisted graphene monolayer on top of the Bernal graphene bilayer. It implies that we can tune the properties of the van der Waals structures efficiently by the stacking structure of their building blocks. Our result may have broad implications in physics of other 2D van der Waals structures.

**Acknowledgements**

We are grateful to National Key Basic Research Program of China (Grant No.

2014CB920903, No. 2013CBA01603), National Science Foundation (Grant No. 11374035, No. 11004010), the program for New Century Excellent Talents in University of the Ministry of Education of China (Grant No. NCET-13-0054), and Beijing Higher Education Young Elite Teacher Project (Grant No. YETP0238).

* Email: helin@bnu.edu.cn